\documentclass[a4paper,11pt]{article}
\pdfoutput=1 
\usepackage{jheppub} 
\usepackage[T1]{fontenc} 
\usepackage{multirow}
\usepackage{color}
\usepackage{ulem}
\usepackage{rotating}
\usepackage{booktabs}
\usepackage{dcolumn}
\usepackage{array} 
\usepackage{multirow}
\usepackage{lineno} 
\usepackage{siunitx}
\usepackage{amsmath}
\usepackage{bm}
\usepackage{mathrsfs}
\usepackage{graphicx}
\usepackage{comment}
\usepackage{float}

\newcommand{\ee}{e^+e^-}
\newcommand{\jpsi}{J/\psi}
\newcommand{\psip}{\psi(3686)}
\newcommand{\psipp}{\psi(3770)}
\newcommand{\piz}{\pi^{0}}
\newcommand{\LLb}{\Lambda\bar{\Lambda}}
\newcommand{\LSb}{\Lambda\bar{\Sigma}^{0}}

\title{\boldmath Partial wave analysis of $\psip\to\LSb\piz+c.c.$}

\collaborationImg{\includegraphics[width=0.25\textwidth]{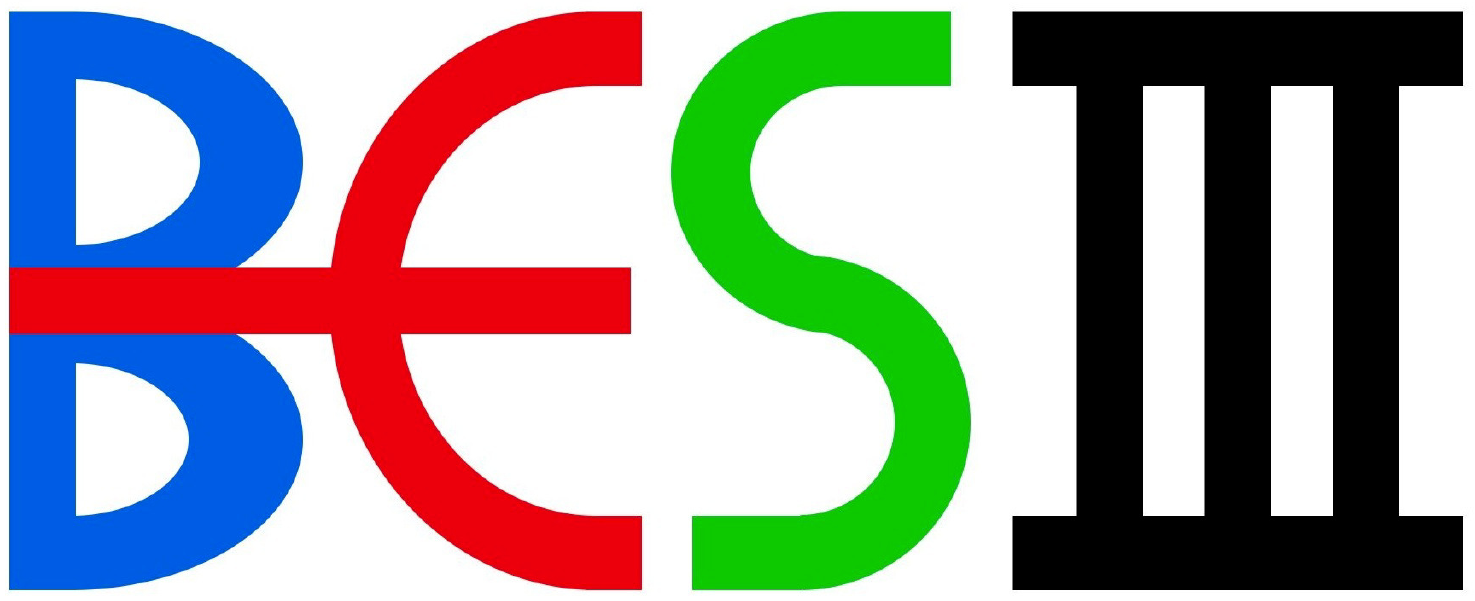}}
\collaboration{The BESIII collaboration}

\emailAdd{besiii-publications@ihep.ac.cn}

\abstract{Based on a sample of $(2712.4\pm14.3)\times10^6\;\psi(3686)$ events collected with the BESIII detector, a partial wave analysis of the decay $\psi(3686)\to\Lambda\bar{\Sigma}^0\piz+c.c.$ is performed to investigate $\Lambda^*$ and $\Sigma^*$ resonances in the $\piz\bar{\Sigma}^0$ and $\piz\Lambda$ invariant mass distributions. Significant contributions are found from the $\Lambda(1405)$, $\Lambda(1520)$, $\Lambda(1600)$, $\Lambda(1670)$, $\Lambda(1690)$, $\Lambda(1800)$, $\Lambda(1890)$, $\Lambda(2325)$, $\Sigma(1385)$, $\Sigma(1660)$, $\Sigma(1670)$, $\Sigma(1750)$, and $\Sigma(1910)$. The masses, widths, and production branching fractions for each component are determined. In addition, the branching fraction of $\psi(3686)\to\Lambda\bar{\Sigma}^0\piz+c.c.$ is measured to be $(1.544\pm0.013\pm0.071)\times10^{-4}$ for the first time, where the first uncertainty is statistical and the second systematic.}

\begin{document}
\maketitle
\flushbottom

\section{Introduction}

The baryons, composed of three quarks, offer the simplest system in which the three colors of quantum-chromodynamics (QCD) neutralize into colorless objects. Therefore, baryon spectroscopy can be used to shed light on QCD. Although the quark model and lattice QCD have achieved significant success in the interpretation of many properties of baryons and their excited states~\cite{Edwards:2011jj}, our current knowledge of baryon spectroscopy is still limited as many fundamental issues are not well understood~\cite{Klempt:2009pi, Wang:2024jyk}. In recent years, there have been many studies of excited states of the nucleons, but there are few results on baryons with a strange quark, namely the $\Lambda$ and $\Sigma$ hyperons~\cite{ParticleDataGroup:2022pth}. In studies of the excited baryons, major difficulties arise from their broad widths and nearby masses, causing a significant amount of overlap between the various excited states~\cite{Wang:2015qta, Liu:2017hdx, He:2024uau}. The partial wave analysis (PWA) technique helps to overcome these difficulties by disentangling different resonances using their spin-parity. In the process, their masses, widths, and partial decay widths can be measured.

The large data samples of $\jpsi$ and $\psip$ events produced from $\ee$ annihilation at BESIII provide excellent opportunities to study different excited baryons. In particular, the $\psip$, which benefits from having a higher mass than the $\jpsi$, has more phase space available in its decays and thus higher mass resonances can be observed. At present, $(2712.4\pm14.3)\times 10^6\;\psip$ events~\cite{BESIII:2024lks} are now available, which allows us to perform extensive studies of baryon spectroscopy. In this paper, a PWA of the decay $\psip\to\LSb\piz+c.c.$ is performed to investigate potential excited states of the $\Lambda$ and $\Sigma$ hyperons. Throughout this paper, the charge-conjugate process is always implied.

\section{BESIII detector and Monte Carlo simulation}

The BESIII detector~\cite{BESIII:2009fln} records symmetric $\ee$ collisions provided by the BEPCII storage ring~\cite{Yu:2016cof} in the center-of-mass energy range from 2.0 to 4.95~GeV, with a peak luminosity of $1.1 \times 10^{33}\;\text{cm}^{-2}\text{s}^{-1}$ achieved at $\sqrt{s} = 3.773\;\text{GeV}$. BESIII has collected large data samples in this energy region~\cite{BESIII:2020nme,Zhang:2022bdc}. The cylindrical core of the BESIII detector covers 93\% of the full solid angle and consists of a helium-based multilayer drift chamber~(MDC), a plastic scintillator time-of-flight system~(TOF), and a CsI(Tl) electromagnetic calorimeter~(EMC), which are all enclosed in a superconducting solenoidal magnet providing a 1.0~T magnetic field. The solenoid is supported by an octagonal flux-return yoke with resistive plate counter muon identification modules interleaved with steel. The charged-particle momentum resolution at $1~{\rm GeV}/c$ is $0.5\%$, and the ${\rm d}E/{\rm d}x$ resolution is $6\%$ for electrons from Bhabha scattering. The EMC measures photon energies with a resolution of $2.5\%$ ($5\%$) at $1$~GeV in the barrel (end cap) region. The time resolution in the TOF barrel region is $68~\mathrm{ps}$, while that in the end cap region was $110~\mathrm{ps}$. The end cap TOF system was upgraded in 2015 using multigap resistive plate chamber technology, providing a time resolution of $60~\mathrm{ps}$, which benefits 83.3\% of the data used in this analysis~\cite{Li:2017jpg,Guo:2017sjt,Cao:2020ibk}.

Simulated data samples produced with a {\sc geant4}-based~\cite{GEANT4:2002zbu} Monte Carlo (MC) package, which includes the geometric description of the BESIII detector and the detector response, are used to determine detection efficiencies and to estimate backgrounds. The simulation models the beam energy spread and initial state radiation (ISR) in the $\ee$ annihilations with the generator {\sc kkmc}~\cite{Jadach:2000ir,Jadach:1999vf}. The inclusive MC sample includes the production of the $\psip$ resonance, the ISR production of the $\jpsi$, and the continuum processes incorporated in {\sc kkmc}~\cite{Jadach:2000ir,Jadach:1999vf}. All particle decays are modelled with {\sc evtgen}~\cite{Lange:2001uf,Ping:2008zz} using branching fractions either taken from the Particle Data Group~(PDG)~\cite{ParticleDataGroup:2022pth}, when available, or otherwise estimated with {\sc lundcharm}~\cite{Chen:2000tv,Yang:2014vra}. Final state radiation from charged final state particles is incorporated using the {\sc photos} package~\cite{Richter-Was:1992hxq}.

\section{Event Selection}

The decay $\psip\to\Lambda\bar{\Sigma}^0\piz$ is reconstructed using the decays $\Sigma^0\to\Lambda\gamma$, $\Lambda\to p\pi^{-}$, and $\pi^{0}\to\gamma\gamma$. Since the final state of the signal decay is $p\bar{p}\pi^+\pi^-\gamma\gamma\gamma$, the number of charged tracks is required to be four with zero net charge. Each charged track must satisfy $\left|\cos\theta\right|<0.93$, where $\theta$ is the polar angle of the track measured by the MDC with respect to the direction of the positron beam. Each of the photon candidates is required to have an energy deposited in the EMC of more than 25 MeV in the barrel $( \left| \cos \theta \right| <0.80)$ or 50 MeV in the end-caps $(0.86< \left| \cos \theta \right| <0.92)$. To eliminate showers from charged tracks, the opening angle between the position of each shower in the EMC and any charged track must be greater than 10 degrees. To suppress electronic noise and showers unrelated to the event, the EMC time difference from the event start time is required to be within $[0,700]~\mathrm{ns}$. At least three photon candidates are required.

\begin{figure}[tbp]
    \centering
    \includegraphics[width=0.485\textwidth]{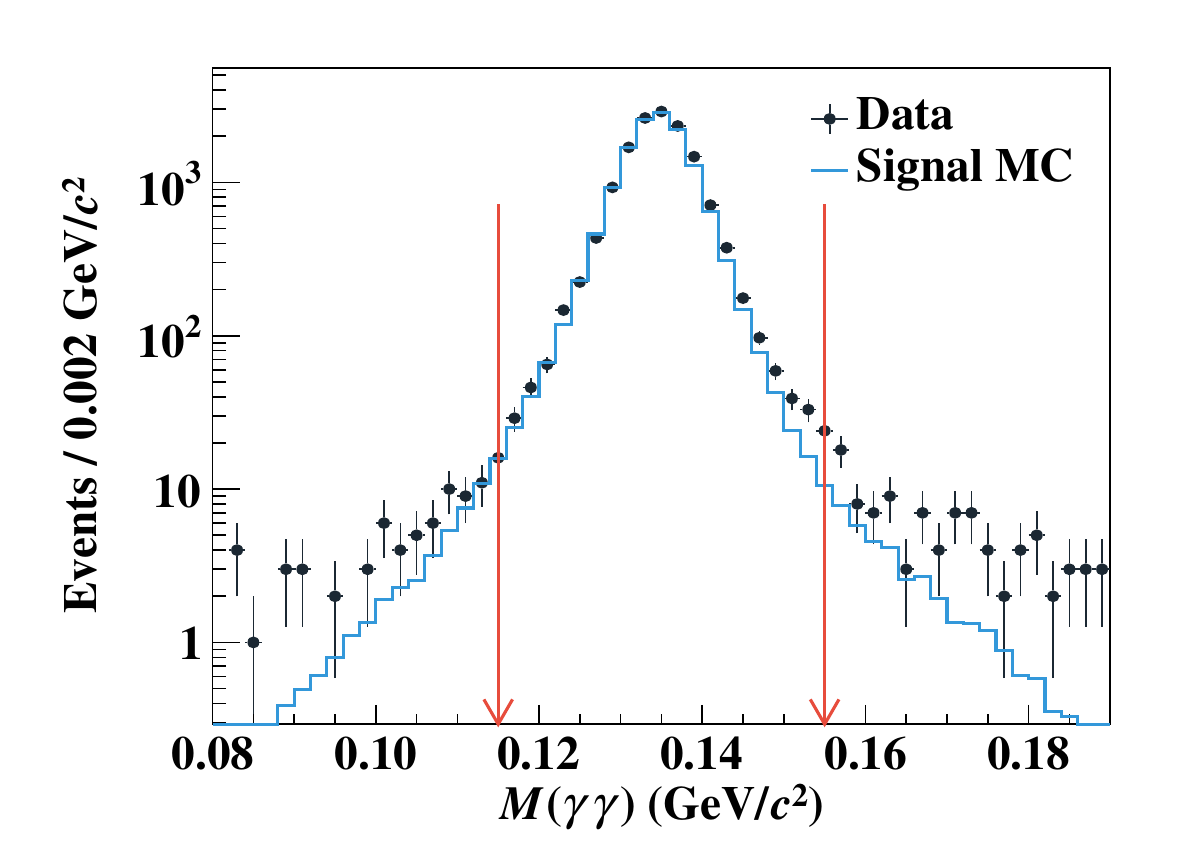}
    \includegraphics[width=0.485\textwidth]{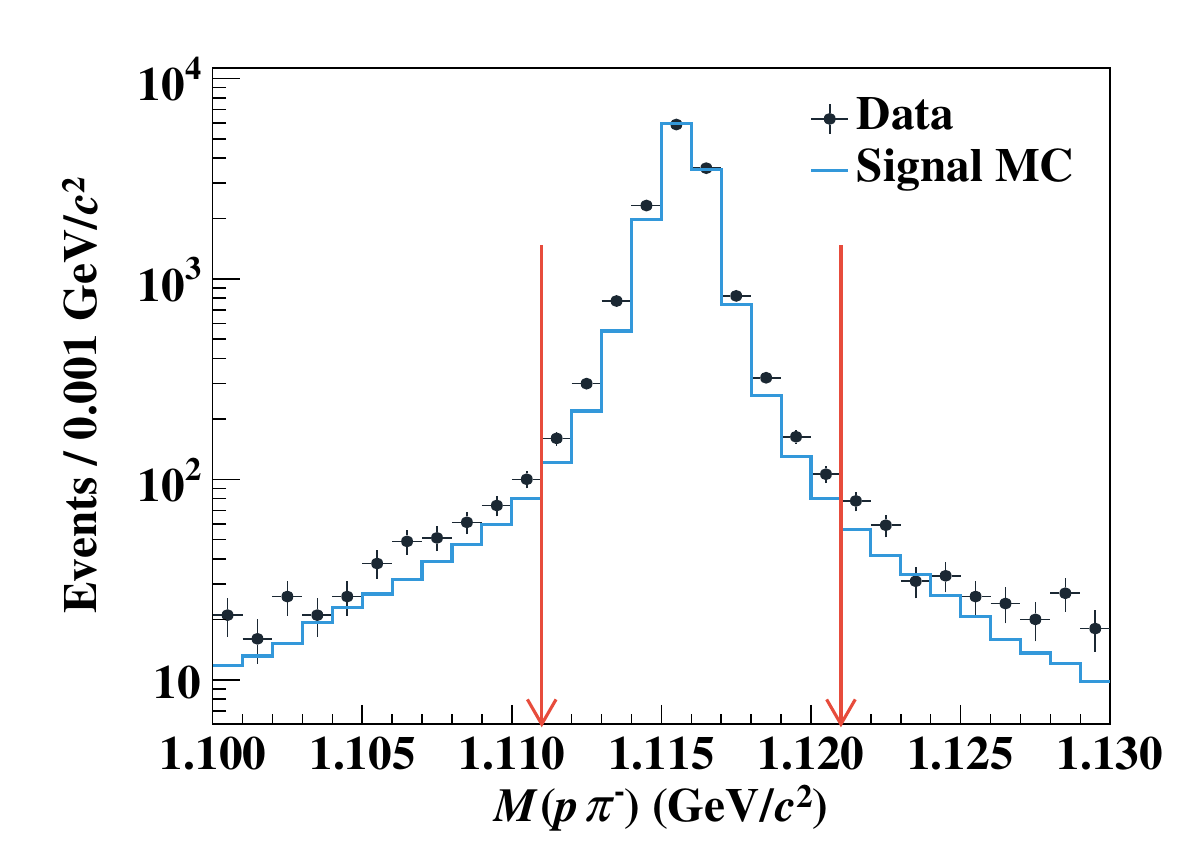}
    \caption{The distributions of $M(\gamma\gamma)$ and $M(p\pi^-)$. The dots with error bars represent data and the blue histograms are the normalized signal MC sample. The signal regions of $\piz$ and $\Lambda$ are shown with the red arrows.}
    \label{fig:line}
\end{figure}

The $\Lambda$ and $\bar{\Lambda}$ candidates are reconstructed by combining pairs of oppositely charged tracks with pion and proton mass hypotheses, fulfilling a secondary vertex constraint~\cite{Xu:2009zzg, BESIII:2019cuv, BESIII:2021ccp, BESIII:2021cvv}. Events with at least one $\Lambda(p\pi^-)$ candidate and one $\bar{\Lambda}(\bar{p}\pi^+)$ candidate are selected. In the case of multiple $\LLb$ pair candidates, the one with the minimum value of $\chi^2_\mathrm{svtx}(\Lambda)+\chi^2_\mathrm{svtx}(\bar{\Lambda})$ is chosen, where $\chi^2_\mathrm{svtx}(\Lambda)$ and $\chi^2_\mathrm{svtx}(\bar{\Lambda})$ are the fit $\chi^2$ of the secondary vertex fits for the $\Lambda$ and $\bar{\Lambda}$, respectively. To improve the momentum and energy resolution and reduce background, a four-constraint (energy-momentum conservation, 4C) kinematic fit is applied under the hypothesis of $\psip\to\LLb\gamma\gamma\gamma$, and the corresponding fit $\chi^2$ ($\chi^2_\mathrm{4C}$) is required to be less than 40. For events with more than three photon candidates, the combination with the least $\chi^2_\mathrm{4C}$ is selected from all possible combinations. To reject possible background events from $\psip\to\LLb\gamma\gamma$ and $\psip\to\LLb\gamma\gamma\gamma\gamma$, we further require that the $\chi^2_\mathrm{4C}$ for the $\psip\to\LLb\gamma\gamma\gamma$ hypothesis is less than those of both the $\psip\to\LLb\gamma\gamma$ and $\psip\to\LLb\gamma\gamma\gamma\gamma$ hypotheses. In the case of multiple $\piz$ candidates, the one with the minimum of $\Delta M(\piz)$ is chosen. The $\Delta M(\piz)$ is defined as $\Delta M(\piz)=\left|M(\gamma_a\gamma_b)-M_\mathrm{PDG}(\piz)\right|$, where $M(\gamma_a\gamma_b)$ is the invariant mass of two photon candidates from the 4C kinematic fit and $M_\mathrm{PDG}(\piz)$ is the nominal $\piz$ mass~\cite{ParticleDataGroup:2022pth}. The final distributions of $M(\gamma\gamma)$ and $M(p\pi^-)$ are shown in figure~\ref{fig:line}, where clear $\piz$ and $\Lambda$ signals are observed. The invariant mass of the chosen $\piz$ candidate is required to be in the $\piz$ signal region, $0.115<M(\piz)<0.155\;\mathrm{GeV}/c^2$; and the invariant mass of $p\pi^-$ is required to be in the $\Lambda$ signal region, $1.111<\mathrm{M}(p\pi^-)<1.121\;\mathrm{GeV}/c^2$.

To suppress mis-combination from the conjugated channel, the invariant mass of $\gamma\Lambda$ from $\psip\to\LSb\piz$ is required to not be in the $\Sigma^0$ signal region $M(\gamma\Lambda)<1.179\;\mathrm{GeV}/c^2$ or $M(\gamma\Lambda)>1.204\;\mathrm{GeV}/c^2$. To reject the peaking background in the $\Sigma^0$ signal region from $\psip\to\gamma\chi_{cJ},\;\chi_{cJ}\to\Sigma^0\bar{\Sigma}^0$, the requirement $\Delta M(\piz\bar{\Sigma}^0)<\Delta M(\Sigma^0\bar{\Sigma}^0)$ is imposed. Here, $\Delta M(\piz\bar{\Sigma}^0)$ is defined as $\Delta M(\piz\bar{\Sigma}^0)=|(M(\piz)-M_\mathrm{PDG}(\piz))|+|(M(\bar{\Sigma}^0)-M_\mathrm{PDG}(\Sigma^0))|$, where $M(\piz)$ and $M(\bar{\Sigma}^0)$ are the invariant masses of the reconstructed $\piz$ and $\bar{\Sigma}^0$, respectively. The $M_\mathrm{PDG}(\Sigma^0)$ is the $\Sigma^0$ nominal mass~\cite{ParticleDataGroup:2022pth}. The $\Delta M(\piz\bar{\Sigma}^0)$ is examined by looping over all three photons to reconstruct the $\piz$, and the remaining photon is used to reconstruct the $\bar{\Sigma}^0$. The $\Delta M(\Sigma^0\bar{\Sigma}^0)$ is defined as $\Delta M(\Sigma^0\bar{\Sigma}^0)=|(M(\Sigma^0)-M_\mathrm{PDG}(\Sigma^0))|+|(M(\bar{\Sigma}^0)-M_\mathrm{PDG}(\Sigma^0))|$, where $M(\Sigma^0)$ and $M(\bar{\Sigma}^0)$ are the invariant masses of the reconstructed $\Sigma^0$ and $\bar{\Sigma}^0$; respectively. The $\Delta M(\Sigma^0\bar{\Sigma}^0)$ is examined by looping over all three photons to reconstruct the $\Sigma^0$ and $\bar{\Sigma}^0$ simultaneously. After applying all the selection criteria mentioned above, around 17,000 events in data remain.

\section{Background study}

To investigate the possible background contributions, the same selection criteria are applied to the inclusive MC sample of 2.7 billion $\psip$ events. An analysis of the surviving events is performed with the generic tool {\sc TopoAna}~\cite{Zhou:2020ksj}. It is found that the background peaking around the $\bar{\Sigma}^0$ nominal mass mainly comes from $\psip\to\gamma\chi_{cJ},\;\chi_{cJ}\to\Sigma^0\bar{\Sigma}^0$, while the mis-combination and other background sources form a flat distribution. The fraction of the background events from $\psip\to\gamma\chi_{cJ},\;\chi_{cJ}\to\Sigma^0\bar{\Sigma}^0$, estimated by using the $\pi^0$ sideband events, is determined to be $0.8\%$.

To estimate the background due to continuum production, the same procedure is performed on the data taken at $\sqrt{s}=3.773\;\mathrm{GeV}$ with an integrated luminosity of $7.9\;\mathrm{fb}^{-1}$~\cite{Ablikim:2013ntc,BESIII:2015equ,BESIII:2024lbn}. The background yield is extracted by an extended unbinned maximum likelihood fit to the $M(\gamma\bar{\Lambda})$ distribution, and normalized to the $\psip$ data taking into account the luminosity and energy-dependent cross section. The normalization factor $f_\mathrm{c}$ is calculated as
\begin{align}
    f_\mathrm{c}=\frac{N_{\psip}^\mathrm{exp}}{N_{\psipp}^\mathrm{obs}}=\frac{\mathcal{L}_{\psip}}{\mathcal{L}_{\psipp}}\cdot\frac{\sigma_{\psip}}{\sigma_{\psipp}}\cdot\frac{\epsilon_{\psip}}{\epsilon_{\psipp}} \label{eq:factor},
\end{align}
where $N_{\psip}^\mathrm{exp}$ and $N_{\psipp}^\mathrm{obs}$ are the number of expected events at the $\psip$ energy point and the number of observed events at the $\psipp$ energy point. The $\mathcal{L}$, $\sigma$, and $\epsilon$ quantities represent the luminosity of data, cross-section, and detection efficiency at each energy point, respectively. The details of the cross-section values can be found in ref.~\cite{Asner:2008nq}. The ratio of detection efficiencies, $\epsilon_{\psip}/\epsilon_{\psipp}$, is determined by MC simulation. The $N_{\psipp}^\mathrm{obs}$ parameter is determined to be $465.9\pm22.5$, where the uncertainty is statistical only. The scale factor is calculated to be $f_\mathrm{c}=0.502\pm0.002$, combining the uncertainties from the cross-section, detection efficiency, and integrated luminosity. After normalization, the background yield from $\ee\to\LSb\piz$ in the $\psip$ data is determined to be $N_{\psip}^\mathrm{exp}=235.2\pm11.3$.

\section{Branching fraction}\label{sec:bf}

\begin{figure}[tbp]
    \centering
    \includegraphics[width=0.8\textwidth]{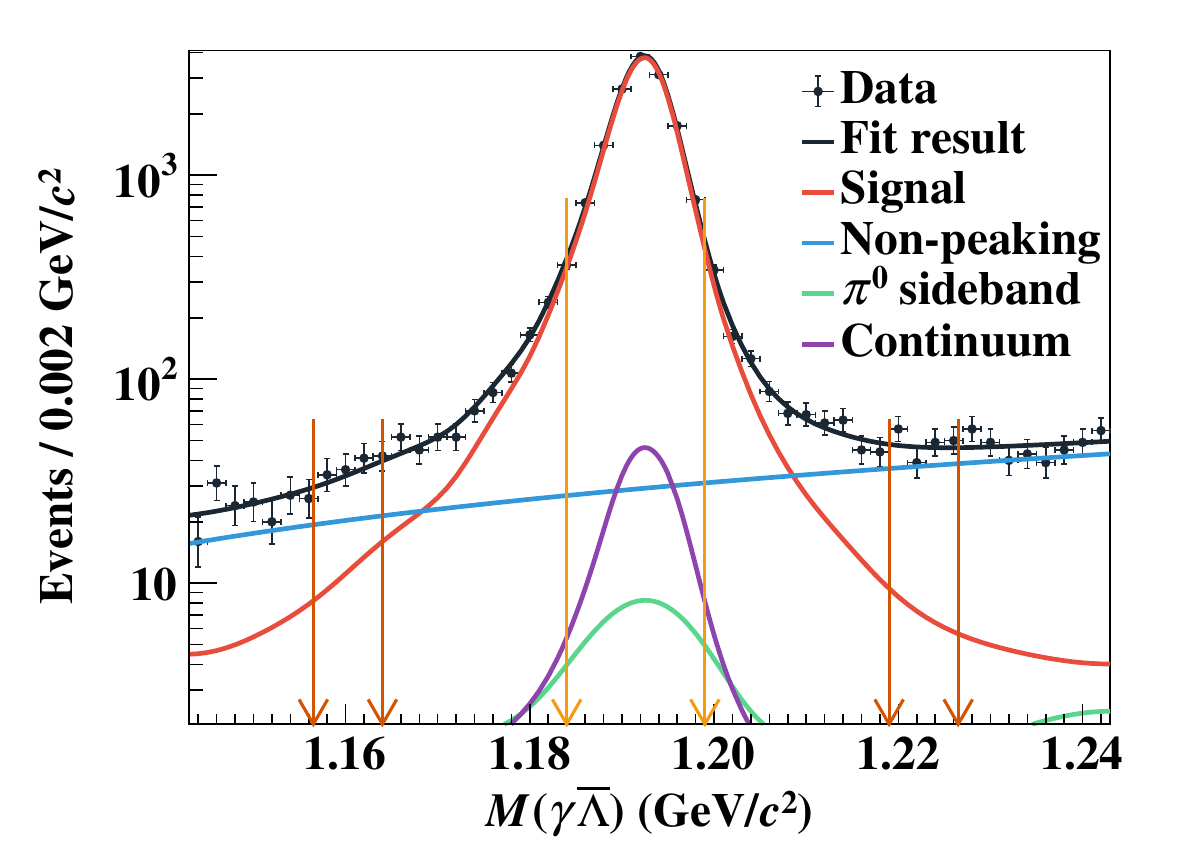}
    \caption{The distribution of $M(\gamma\bar{\Lambda})$ with the result of the fit overlaid. The $\bar{\Sigma}^0$ signal region is shown with the high yellow arrows and the $\bar{\Sigma}^0$ sideband ranges are shown with the short orange arrows.}
    \label{fig:fit_m}
\end{figure}

To determine the signal yield, an extended unbinned maximum likelihood fit is performed on the $M(\gamma\bar{\Lambda})$ distribution. In the fit, the $\bar{\Sigma}^0$ is described by the signal MC shape convolved with a Gaussian function. The non-peaking and mis-combination backgrounds are parameterized by a first-order Chebychev polynomial function. The continuum background shape is derived from the data taken at $\sqrt{s}=3.773\;\mathrm{GeV}$. The peaking background from $\chi_{cJ}\to\Sigma^0\bar{\Sigma}^0$ is estimated by the $\piz$ sideband defined as $M(\gamma\gamma)\in(0.08,0.1)\;\mathrm{GeV}/c^2$ or $(0.17,0.19)\;\mathrm{GeV}/c^2$. The number of $\piz$ sideband events is estimated to be $N_{\piz}^{\mathrm{side}}=126.0\pm11.2$, where the uncertainty is statistical only. The fit results are shown in figure~\ref{fig:fit_m}, from which we obtain $15560.9\pm131.2$ signal events.

With the detection efficiency obtained from the signal MC sample generated with the PWA results, the branching fraction of $\psip\to\Lambda\bar{\Sigma}^0\piz$ is determined from
\begin{align}
    \mathcal{B}(\psip\to\Lambda\bar{\Sigma}^0\piz+c.c.)=\frac{N^\mathrm{obs}}{N_{\psip}\cdot\mathcal{B}^2(\Lambda\to p\pi^-)\cdot\mathcal{B}(\piz\to\gamma\gamma)\cdot\mathcal{B}(\Sigma^0\to\gamma\Lambda)\cdot\epsilon}
\end{align}
where $N^\mathrm{obs}$ is the number of signal events in data, $N_{\psip}$ is the number of $\psip$ events in data determined with inclusive hadronic events~\cite{BESIII:2024lks}, $\epsilon$ is the detection efficiency obtained to be $9.15\%$ from MC simulation, and $\mathcal{B}(\Lambda\to p\pi^-)$, $\mathcal{B}(\piz\to\gamma\gamma)$, and $\mathcal{B}(\Sigma^0\to\gamma\Lambda)$ are the corresponding branching fractions quoted from the PDG~\cite{ParticleDataGroup:2022pth}. The branching fraction $\mathcal{B}(\psip\to\Lambda\bar{\Sigma}^0\piz+c.c.)$ is measured to be $(1.544\pm0.013\pm0.071)\times10^{-4}$, where the first uncertainty is statistical and the second systematic. The systematic uncertainty estimation is detailed in section~\ref{sec:syu}.

\section{Partial wave analysis}

Figure~\ref{fig:fit_m} presents the reconstructed mass distribution of the $\gamma\bar{\Lambda}$ system. To improve the signal purity, an additional requirement on the $\gamma\bar{\Lambda}$ invariant mass, $1.184\;\mathrm{GeV}/c^2<M(\gamma\bar{\Lambda})<1.199\;\mathrm{GeV}/c^2$, is applied. After applying all the selection criteria mentioned above, 14,414 events from the $\psip$ data are selected for the PWA. The $\bar{\Sigma}^0$ sidebands (defined as $M(\gamma\bar{\Lambda})$ within $(1.157,1.164)\;\mathrm{GeV}/c^2$ or $(1.219,1.226)\;\mathrm{GeV}/c^2$) and the data at $\sqrt{s}=3.773\;\mathrm{GeV}$ are used to estimate the yields of the mis-combination and continuum background events, respectively. As the contribution from non-$\piz$ background is small, the $\piz$ sidebands are excluded from the PWA. The numbers of $\bar{\Sigma}^0$ sideband and continuum background events are estimated to be $N^{\mathrm{side}}_{\bar{\Sigma}^0}=320.0\pm17.9$ and $N^\mathrm{cont.}=208.9\pm10.1$, respectively. Here, the uncertainties are statistical only. The number of $\bar{\Sigma}^0$ sideband events is fixed with the normalization factor $1.0$. Finally, the signal purity is estimated to be $96.3\%$. The background-subtraction procedure in the PWA is implemented within the negative log-likelihood ($NLL$) value, which can be expressed as follows:
\begin{align}
    NLL                 & =NLL(\mathrm{data})-\sum_i NLL(\mathrm{bkg}_i),                      \\
    NLL(\mathrm{bkg}_i) & =\sum_j\left(NLL_j(\mathrm{bkg}_i)\times W_j(\mathrm{bkg}_i)\right),
\end{align}
where $NLL(\mathrm{data})$ represents the $NLL$ value of the observed data, and $NLL(\mathrm{bkg}_i)$ denotes the $NLL$ value for the $i^{th}$ background category. The background samples used in the $NLL(\mathrm{bkg}_i)$ calculation are those from which their respective yields in the signal region are derived. The terms $NLL_j(\mathrm{bkg}_i)$ and $W_j(\mathrm{bkg}_i)$ correspond to the $NLL$ value and normalization factor associated with the $j^{th}$ background events of the $i^{th}$ background category, respectively. No mass constraint is applied for the final state particles $\Lambda$, $\bar{\Sigma}^0$, and $\piz$ in the PWA. The decay amplitude is constructed using the helicity amplitude formalism, and the full procedure is implemented based on the open-source framework TF-PWA~\cite{Jiang:2023tf}.

\subsection{Helicity formalism}

To construct the full decay amplitude of $\psip\to\LSb\piz$, the helicity formalism is used in conjunction with the isobar model, where the three-body decay is described as a two-step sequential quasi-two-body decay. For each two-body decay $0\to1+2$, the helicity amplitude can be written as
\begin{align}
    A^{0 \rightarrow 1+2}_{\lambda_{0},\lambda_{1},\lambda_{2}} = H_{\lambda_{1},\lambda_{2}}^{0 \rightarrow 1+2} D^{J_{0^*}}_{\lambda_{0},\lambda_{1}-\lambda_{2}}(\phi,\theta,0),
\end{align}
where the amplitude $H_{\lambda_{1},\lambda_{2}}^{0 \rightarrow 1+2}$ is given by the LS coupling formula~\cite{Chung:1997jn} along with the barrier factor terms
\begin{align}
    H_{\lambda_{1},\lambda_{2}}^{0 \rightarrow 1+2} =
    \sum_{ls} g_{ls} \sqrt{\frac{2l+1}{2 J_{0}+1}} \langle l 0; s \delta|J_{0} \delta\rangle \langle J_{1} \lambda_{1} ;J_{2} -\lambda_{2} | s \delta \rangle q^{l} B_{l}'(q, q_0, d),
\end{align}
where $l$ is the orbital angular momentum, $g_{ls}$ are the fit parameters, $J_{0,1,2}$ are the spins of particles 0, 1, and 2, $\lambda_{1,2}$ are the helicities for particles 1 and 2, and $\delta=\lambda_1-\lambda_2$ is the helicity difference. Here, $q$ is the three-momentum modulus of particle 1 in the rest frame of particle 0, which is calculated as
\begin{align}
    q=\frac{\sqrt{[m^2-(m_1+m_2)^2][m^2-(m_1-m_2)^2]}}{2m},
\end{align}
where $m$, $m_1$, and $m_2$ are the masses of particles 0, 1, and 2, respectively. The normalization factor $q_0$ is calculated at the nominal resonance mass. The factor $B_{l}'(q, q_0, d)$ is the reduced Blatt-Weisskopf barrier factor~\cite{Chung:1993da}. In the Wigner $D$-function, $D^{J_{0}*}_{\lambda_0,\lambda_1 - \lambda_2}(\phi, \theta, 0)$, $\phi$ and $\theta$ are the helicity angles. The definitions can be found in ref.~\cite{Wang:2020giv}. The radius $d$ is chosen as $d=0.73\;\mathrm{fm}$ following ref.~\cite{BESIII:2019dme}.

The amplitude for a complete decay chain is constructed as the product of each two body decay amplitude and the resonant propagator $R$. For example, in the sequential decay $\psip\to\Sigma^{*0}\bar{\Sigma}^0,\;\Sigma^{*0}\to\Lambda\piz$, the amplitude is written as
\begin{align}
    A^{\psip\to\Sigma^{*0}\bar{\Sigma}^0,\;\Sigma^{*0}\to\Lambda\piz}_{\lambda_{\psip},\lambda_{\Lambda},\lambda_{\bar{\Sigma}^0},\lambda_{\piz}}
    = \sum_{\lambda_{\Sigma^{*0}}}A^{\psip\to\Sigma^{*0}\bar{\Sigma}^0}_{\lambda_{\psip},\lambda_{\Sigma^{*0}},\lambda_{\bar{\Sigma}^0}} R(m_{\Sigma^{*0}}) A^{\Sigma^{*0}\to\Lambda\piz} _{\lambda_{\Sigma^{*0}},\lambda_{\Lambda},\lambda_{\piz}}.
\end{align}
The propagator $R$ includes different models. For all resonances except $\Lambda(1405)$, the relativistic Breit-Wigner formula with mass dependent width is taken as
\begin{align}
    R(m)\equiv\frac{1}{m_0^2-s-i\sum_jg_j^2\rho_j(m)}\label{eq:bw},
\end{align}
which iterates over the $j$ decay channels that the resonance may decay into with coupling $g_j$ and phase-space factor $\rho_j$. Then in the case where only decays into the final state are being considered, the imaginary part reduces to
\begin{align}
    \sum_jg_j^2\rho_j(m)=m_0\Gamma(m)=m_0\Gamma_0\left(\frac{q}{q_0}\right)^{2l+1}\frac{m_0}{m}B_l'^2(q,q_0,d)
\end{align}
where $\Gamma_0$ is the width of the resonance, and $q$ and $q_0$ are the momenta in the $m$ and $m_0$ center-mass-system. The radius $d$ is also chosen as $d=0.73\;\mathrm{fm}$ following ref.~\cite{BESIII:2019dme}.

We describe the two-pole structure of the $\Lambda(1405)$ using the chiral dynamics model~\cite{Jido:2003cb}, which is parameterized as
\begin{align}
    R(m)=       & \frac{1}{\left| I-VG \right|},                                                                                   \\
    V_{ij}(m)=  & -C_{ij}\frac{1}{4f^2}(2m-M_i-M_j)\sqrt{\frac{M_i-E_i}{M_i}}\sqrt{\frac{M_j-E_j}{M_j}},                           \\
    G_k(m;\mu)= & \frac{1}{(4\pi)^2}\bigg\{a_k(\mu)+\ln\frac{M_k^2}{\mu}+\frac{m_k^2-M_k^2+m^2}{2m^2}\ln\frac{m_k^2}{M_k^2} \notag \\
                & +\frac{q_k}{m}\big[\ln(m^2-(M_k^2-m_k^2)+2q_km)+\ln(m^2+(M_k^2-m_k^2)+2q_km)                              \notag \\
                & -\ln(-m^2+(M_k^2-m_k^2)+2q_km)-\ln(-m^2-(M_k^2-m_k^2)+2q_km)\big]\bigg\},
\end{align}
\noindent where $I$ is the unit matrix, $m=\sqrt{s}$ is the mass of the event, the $C_{ij}$ coefficients are given in ref.~\cite{Oset:1997it}, an averaged meson decay constant $f=1.123f_\pi$ is used~\cite{Oset:2001cn}, with $f_\pi=92.4\;\mathrm{MeV}$ being the weak pion decay constant, and $M_i$ and $E_i$ are the physical mass and energy of the baryon. The $\mu$ is the scale of dimensional regularization fixed to be $630\;\mathrm{MeV}$~\cite{Oset:2001cn}, the subtraction constant $a_k(\mu)$ is quoted from ref.~\cite{Oset:2001cn}, $m_i$ is the physical mass of the meson, and $q_k$ is the momentum in the rest frame of $\Lambda(1405)$.

The construction of the probability density function, the calculation of the fit fraction, and the corresponding statistical uncertainty for each component follow ref.~\cite{BESIII:2022udq}. To consider the effect of mass resolution, we use the method from the TF-PWA documentation~\cite{Jiang:2023tf}.

\subsection{Nominal fit model}

To determine the nominal fit hypothesis of the PWA, the statistical significance of each potential component is evaluated based on the change of the $NLL$ value when that component is included, taking into account the change of the number of degrees of freedom ($N_\mathrm{dof}$). The components considered are excited $\Lambda$ and $\Sigma$ states, including the $\Lambda(1405)$, $\Lambda(1520)$, $\Lambda(1600)$, $\Lambda(1670)$, $\Lambda(1690)$, $\Lambda(1800)$, $\Lambda(1810)$, $\Lambda(1890)$, $\Lambda(2325)$, $\Sigma(1385)$, $\Sigma(1660)$, $\Sigma(1670)$, $\Sigma(1750)$, and $\Sigma(1910)$ (considered as established by the PDG~\cite{ParticleDataGroup:2022pth} with a status of at least three-star of existence~\footnote{The overall states for existence of baryon resonances~\cite{ParticleDataGroup:2022pth}: ****: Existence is certain; ***: Existence is very likely; **: Evidence of existence is fair; *: Evidence of existence is poor.} and a spin less than $5/2$), as well as the $\rho_3(2250)$, $\rho_5(2350)$, and $\mathcal{S}$-wave non-resonant component ($NR_{1^-}$) in the $M(\Lambda\bar{\Sigma}^0)$ distribution. In addition, an extra $\Lambda$ state, the $\Lambda(2325)$, is taken into account to describe the high side of the $M(\piz\bar{\Sigma}^0)$ distribution.  The results show that the resonances $\Lambda(1405)$, $\Lambda(1520)$, $\Lambda(1600)$, $\Lambda(1670)$, $\Lambda(1690)$, $\Lambda(1800)$, $\Lambda(1890)$, $\Lambda(2325)$, $\Sigma(1385)$, $\Sigma(1660)$, $\Sigma(1670)$, $\Sigma(1750)$, and $\Sigma(1910)$ have statistical significance greater than $5\sigma$, while none of the other tested contributions exceed this threshold.

Based on the MC simulation, the mass resolution for the $M(\piz\bar{\Sigma}^0)$ distribution is estimated to be $4.8\;\mathrm{MeV}$. The widths of the $\Lambda(1520)$ and $\Lambda(1670)$ states from the PDG~\cite{ParticleDataGroup:2022pth} are $15\sim17\;\mathrm{MeV}$ and $25\sim35\;\mathrm{MeV}$, respectively, which are close to the mass resolution. Therefore, the mass resolution on the $M(\piz\bar{\Sigma}^0)$ distribution is taken into account in the nominal fit hypothesis. In the fit, the masses and widths of all resonances are left as free parameters.

A resonance of one-star of existence, the $\Lambda(2325)$, is also added to the nominal fit. It was first reported in ref.~\cite{Baccari:1976ik} with $J^P=3/2^-$. A check on the spin-parity of the $\Lambda(2325)$ has been performed. A comparison of the statistical significance with the different spin-parity hypotheses ($1/2^-$, $1/2^+$, $3/2^-$ and $3/2^+$) is shown in table~\ref{tab:JP_scan_L2325}. The statistical significance is calculated based on the change of the $NLL$ values with and without including the component, by taking into account the change of the number of degrees of freedom. Based on the comparison, the spin-parity of the $\Lambda(2325)$ is most likely $J^P=3/2^-$, which is consistent with the PDG~\cite{ParticleDataGroup:2022pth}.

\begin{table}[tbp]
    \centering
    \begin{tabular}{cccc}
        \toprule
        $J^P$   & $\Delta NLL$ & $\Delta N_\mathrm{dof}$ & Significance ($\sigma$) \\
        \midrule
        $1/2^-$ & 188.0        & 6                       & 18.7                    \\
        $1/2^+$ & 186.8        & 6                       & 18.6                    \\
        $3/2^-$ & 208.3        & 8                       & 19.5                    \\
        $3/2^+$ & 198.8        & 8                       & 19.1                    \\
        \bottomrule
    \end{tabular}
    \caption{Comparison of the different quantum numbers for the spin-parity of the $\Lambda(2325)$.}
    \label{tab:JP_scan_L2325}
\end{table}

The branching fractions of the intermediate states $i$, are determined by
\begin{align}
    \mathcal{B}_i=\mathcal{B}(\psi(3686)\to\Lambda\bar{\Sigma}^0\pi^0+c.c.)FF_i,
\end{align}
where $FF_i$ is the fit fraction of the $i^{th}$ intermediate state.

\section{Systematic uncertainties}\label{sec:syu}

\subsection{Systematic uncertainties on branching fraction}

The systematic uncertainties on the branching fraction measurement as listed in table~\ref{tab:br_syu} are discussed in detail below. Assuming all sources are independent, the total systematic uncertainty is determined by adding them in quadrature.

\begin{enumerate}
    \item The efficiency of the $\Lambda(\bar{\Lambda})$ reconstruction, incorporating both the MDC tracking and the $\Lambda(\bar{\Lambda})$ mass window requirement, is studied using a control sample of $\psip\to\Lambda\bar{\Lambda}$ decays, and a correction factor of $0.980\pm0.011$~\cite{Ablikim:2017tys} is applied to the detection efficiencies obtained from MC simulation.
    \item For photons directly detected by the EMC, the detection efficiency is studied using a control sample of $\ee\to\gamma_\mathrm{ISR}\mu^+\mu^-$, where ISR stands for initial state radiation. The systematic uncertainty, defined as the relative difference in efficiencies between data and MC simulation, is observed to be up to $0.5\%$ per photon in both the barrel and end-cap regions. Thus, the total systematic uncertainty of the three photons is $1.5\%$.
    \item The uncertainty associated with the 4C kinematic fit is due to the inconsistency between data and MC simulation. This difference has been reduced by correcting the track helix parameters in the MC simulation, with parameters taken from refs.~\cite{BESIII:2013nam, BESIII:2019gjc}. Following the method described in ref.~\cite{Ablikim:2012pg}, the difference between the corrected and uncorrected efficiencies, $1.4\%$, is assigned as the systematic uncertainty due to the 4C kinematic fit.
    \item The uncertainty associated with the $\piz$ signal region is assessed by evaluating the proportional impact the constraint has on the number of $\piz$ candidates remaining in both the data and signal MC samples. The difference in the branching fraction, $1.2\%$, is assigned as the systematic uncertainty due to the $\piz$ signal region.
    \item In order to evaluate the systematic uncertainty due to the signal shape, an alternative fit is performed to determine the number of signal events. The simulated signal shape is replaced with a Crystal-ball function convolved with a Gaussian function. The maximum difference in branching fraction, $2.5\%$, is assigned as the corresponding systematic uncertainty.
    \item To estimate the uncertainty of the shape of the non-peaking and mis-combination, $\pi^0$ sideband and continuum backgrounds, we perform alternative fits by replacing the shape of each background. The first-order Chebychev polynomial function is replaced with a second-order Chebychev polynomial function for the non-peaking and mis-combination background. The shape of the $\pi^0$ sideband and continuum background is replaced with a Crystal-ball function convolved with a Gaussian function, whose parameters are extracted from the fit to data. The uncertainty associated with the normalization factor \( f_\mathrm{c} \) for the continuum background is estimated to be less than \( 0.1\% \), which can be considered negligible. The difference in branching fraction is combined to be $0.4\%$, which is assigned as the systematic uncertainty due to the background shape.
    \item The uncertainty due to the MC model is evaluated by comparing the efficiencies between the nominal and alternative amplitude models. The efficiency difference between different models is taken as the systematic uncertainty.
    \item The uncertainties from the quoted branching fractions of the intermediate decays of $\Lambda\to p\pi^-$ and $\piz\to\gamma\gamma$ are taken from the PDG~\cite{ParticleDataGroup:2022pth}. Since the branching fraction of $\Sigma^0\to\gamma\Lambda$ is equal to $100\%$ with no uncertainty, it is not considered in the systematic uncertainty.
    \item The total number of $\psip$ events in data is determined using inclusive hadronic decays. Its uncertainty, $0.5\%$~\cite{BESIII:2024lks}, is taken as a systematic uncertainty.
\end{enumerate}

\begin{table}[tbp]
    \centering
    \begin{tabular}{l|c}
        \toprule
        Source                                  & Uncertainty~$(\%)$ \\
        \midrule
        $\Lambda(\bar{\Lambda})$ reconstruction & 2.2                \\
        Photon reconstruction                   & 1.5                \\
        Kinematic fit                           & 1.4                \\
        $\piz$ signal region                    & 1.2                \\
        Signal shape                            & 2.5                \\
        Background shape                        & 0.4                \\
        MC model                                & 1.1                \\
        Quoted branching fractions              & 1.6                \\
        Total number of $\psip$ events          & 0.5                \\
        \hline
        Total                                   & 4.6                \\
        \bottomrule
    \end{tabular}
    \caption{The systematic uncertainties for the branching fraction.}
    \label{tab:br_syu}
\end{table}

\subsection{Systematic uncertainties on partial wave analysis}

The systematic uncertainties on the PWA as listed in tables~\ref{tab:pwa_syu_mass},~\ref{tab:pwa_syu_width}, and~\ref{tab:pwa_syu_ff} are discussed in detail below. Assuming all sources are independent, the total systematic uncertainty is determined by adding them in quadrature.

\begin{enumerate}
    \item To evaluate the effect on the PWA results from other possible components (EX-RES), the PWA is re-performed by adding extra resonances ($\Lambda(1810)$, $\Lambda(1820)$, $\Lambda(1830)$, $\Sigma(1775)$, $\Sigma(1915)$, $\rho_3(2250)$, $\rho_5(2350)$) one at a time. The largest changes of the masses, widths, and fit fractions of resonances are taken as the systematic uncertainties.
    \item The uncertainty associated with the background estimation (BKG) is assessed using two distinct methods. The first method involves varying the yields of the $\bar{\Sigma}^0$ sideband and continuum production within $\pm 1\sigma$ of their statistical uncertainties. The second method incorporates the $\piz$ sidebands, as defined in Sec.~\ref{sec:bf}, to estimate the background while fixing the yields of both sidebands with a normalization factor of $0.5$. The largest difference observed for each result is considered as the systematic uncertainty.
    \item The uncertainty due to the $\Lambda(1405)$ parameterization (PARAM) is evaluated by replacing the nominal formula by the Flatt\'{e}-like formula. The differences of the masses, widths, and fit fractions of resonances are taken as the systematic uncertainties.
    \item In the PWA, the radius $d$ in the Blatt-Weisskopf barrier factors~\cite{Chung:1993da} (BW-BF) is chosen as $d=0.73\;\mathrm{fm}\approx3.7\;\mathrm{GeV}^{-1}$ following ref.~\cite{BESIII:2019dme}. The associated systematic uncertainty is evaluated by varying the radius $d$ in the range $d\in[1,5]\;\mathrm{GeV}^{-1}$~\cite{ParticleDataGroup:2022pth}. The largest differences of the masses, widths, and fit fractions of resonances are taken as the systematic uncertainties.
\end{enumerate}

\begin{table}[tbp]
    \centering
    \begin{tabular}{c|rrrr|r}
        \toprule
        Resonance       & EX-RES & BKG    & PARAM & BW-BF  & Total  \\
        \midrule
        $\Lambda(1520)$ & $4.1$  & $2.4$  & $0.1$ & $0.5$  & $4.8$  \\
        $\Lambda(1600)$ & $8.8$  & $6.4$  & $2.8$ & $4.2$  & $12.1$ \\
        $\Lambda(1670)$ & $1.2$  & $3.2$  & $0.6$ & $0.8$  & $3.6$  \\
        $\Lambda(1690)$ & $7.7$  & $13.3$ & $2.2$ & $2.2$  & $15.7$ \\
        $\Lambda(1800)$ & $2.2$  & $2.9$  & $8.6$ & $8.5$  & $12.6$ \\
        $\Lambda(1890)$ & $1.9$  & $2.6$  & $2.1$ & $4.9$  & $6.2$  \\
        $\Lambda(2325)$ & $10.4$ & $10.4$ & $1.9$ & $8.5$  & $17.1$ \\
        $\Sigma(1385)$  & $1.5$  & $2.0$  & $0.3$ & $2.1$  & $3.3$  \\
        $\Sigma(1660)$  & $4.7$  & $5.4$  & $2.6$ & $0.5$  & $7.6$  \\
        $\Sigma(1670)$  & $1.8$  & $2.3$  & $3.1$ & $0.5$  & $4.3$  \\
        $\Sigma(1750)$  & $6.0$  & $2.0$  & $1.0$ & $3.5$  & $7.3$  \\
        $\Sigma(1910)$  & $23.7$ & $5.8$  & $9.8$ & $21.0$ & $33.6$ \\
        \bottomrule
    \end{tabular}
    \caption{The systematic uncertainty sources for the masses in units of $\mathrm{MeV}/c^2$.}
    \label{tab:pwa_syu_mass}
\end{table}

\begin{table}[tbp]
    \centering
    \begin{tabular}{c|rrrr|r}
        \toprule
        Resonance       & EX-RES & BKG    & PARAM  & BW-BF  & Total  \\
        \midrule
        $\Lambda(1520)$ & $0.2$  & $0.3$  & $0.3$  & $0.4$  & $0.6$  \\
        $\Lambda(1600)$ & $30.3$ & $12.2$ & $9.1$  & $8.2$  & $34.8$ \\
        $\Lambda(1670)$ & $5.0$  & $6.7$  & $0.1$  & $2.5$  & $8.7$  \\
        $\Lambda(1690)$ & $9.8$  & $15.3$ & $1.7$  & $0.9$  & $18.3$ \\
        $\Lambda(1800)$ & $35.2$ & $4.1$  & $3.7$  & $9.0$  & $36.7$ \\
        $\Lambda(1890)$ & $2.0$  & $2.7$  & $1.7$  & $1.6$  & $4.1$  \\
        $\Lambda(2325)$ & $47.5$ & $2.9$  & $1.7$  & $4.1$  & $47.8$ \\
        $\Sigma(1385)$  & $5.2$  & $0.6$  & $0.8$  & $3.4$  & $6.3$  \\
        $\Sigma(1660)$  & $39.2$ & $5.1$  & $<0.1$ & $11.6$ & $41.1$ \\
        $\Sigma(1670)$  & $4.5$  & $6.0$  & $3.7$  & $2.9$  & $8.9$  \\
        $\Sigma(1750)$  & $8.3$  & $3.7$  & $2.5$  & $2.2$  & $9.7$  \\
        $\Sigma(1910)$  & $37.5$ & $23.7$ & $5.5$  & $10.6$ & $46.0$ \\
        \bottomrule
    \end{tabular}
    \caption{The systematic uncertainty sources for the widths in units of $\mathrm{MeV}$.}
    \label{tab:pwa_syu_width}
\end{table}

\begin{table}[tbp]
    \centering
    \begin{tabular}{c|rrrr|r}
        \toprule
        Resonance       & EX-RES & BKG    & PARAM  & BW-BF & Total  \\
        \midrule
        $\Lambda(1405)$ & $21.0$ & $8.9$  & $11.8$ & $0.8$ & $25.7$ \\
        $\Lambda(1520)$ & $19.9$ & $7.9$  & $12.9$ & $8.3$ & $26.3$ \\
        $\Lambda(1600)$ & $21.2$ & $5.0$  & $15.3$ & $4.5$ & $27.0$ \\
        $\Lambda(1670)$ & $21.9$ & $16.9$ & $0.7$  & $5.7$ & $28.3$ \\
        $\Lambda(1690)$ & $14.3$ & $8.0$  & $4.9$  & $5.9$ & $18.1$ \\
        $\Lambda(1800)$ & $24.2$ & $15.2$ & $3.4$  & $8.3$ & $30.0$ \\
        $\Lambda(1890)$ & $8.5$  & $13.6$ & $2.3$  & $7.4$ & $17.8$ \\
        $\Lambda(2325)$ & $2.5$  & $5.9$  & $2.4$  & $3.2$ & $7.6$  \\
        $\Sigma(1385)$  & $11.7$ & $4.8$  & $4.5$  & $4.9$ & $14.2$ \\
        $\Sigma(1660)$  & $13.6$ & $7.8$  & $5.3$  & $4.2$ & $17.1$ \\
        $\Sigma(1670)$  & $17.5$ & $8.3$  & $11.9$ & $4.9$ & $23.3$ \\
        $\Sigma(1750)$  & $19.4$ & $5.4$  & $6.4$  & $8.9$ & $22.9$ \\
        $\Sigma(1910)$  & $22.0$ & $3.9$  & $9.9$  & $9.9$ & $26.3$ \\
        \bottomrule
    \end{tabular}
    \caption{The systematic uncertainty sources for the fit fractions in percentage.}
    \label{tab:pwa_syu_ff}
\end{table}

\section{Results and interpretation}

\begin{figure}[tbp]
    \centering
    \includegraphics[width=0.485\textwidth]{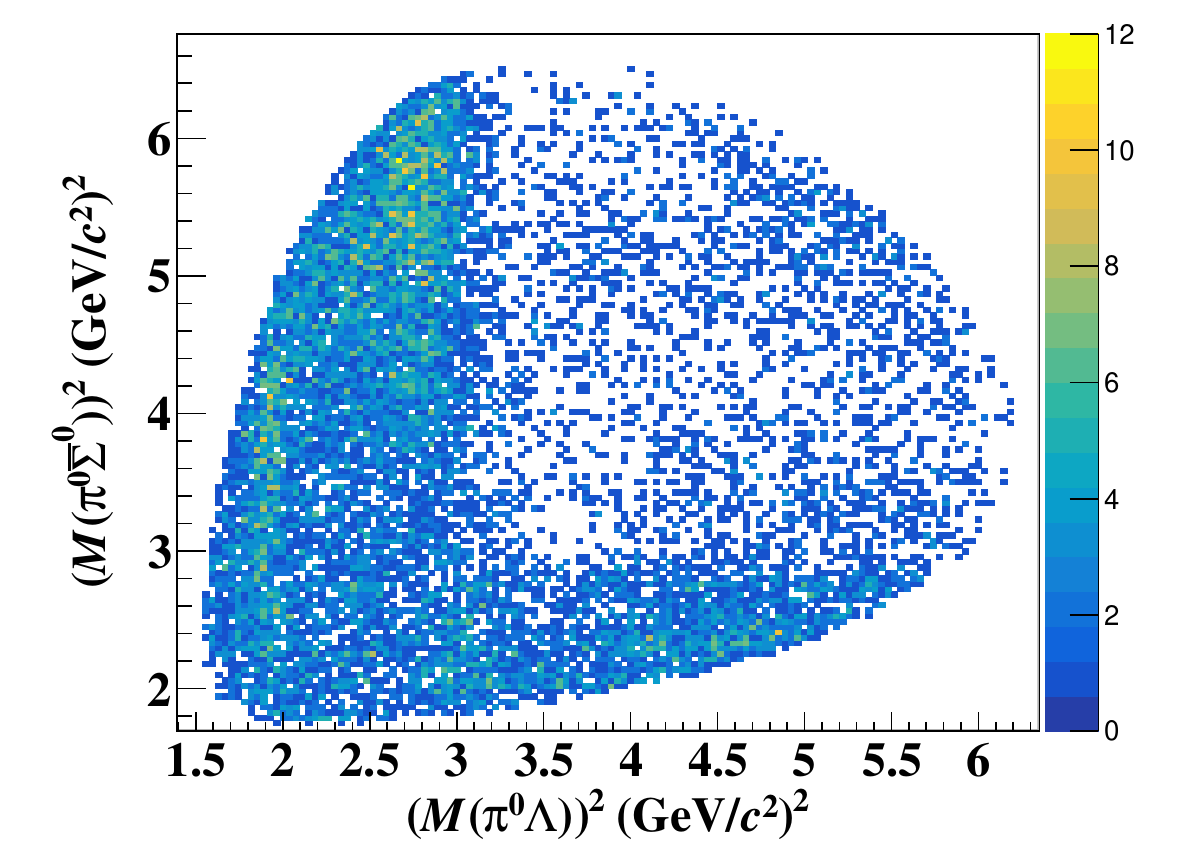}
    \includegraphics[width=0.485\textwidth]{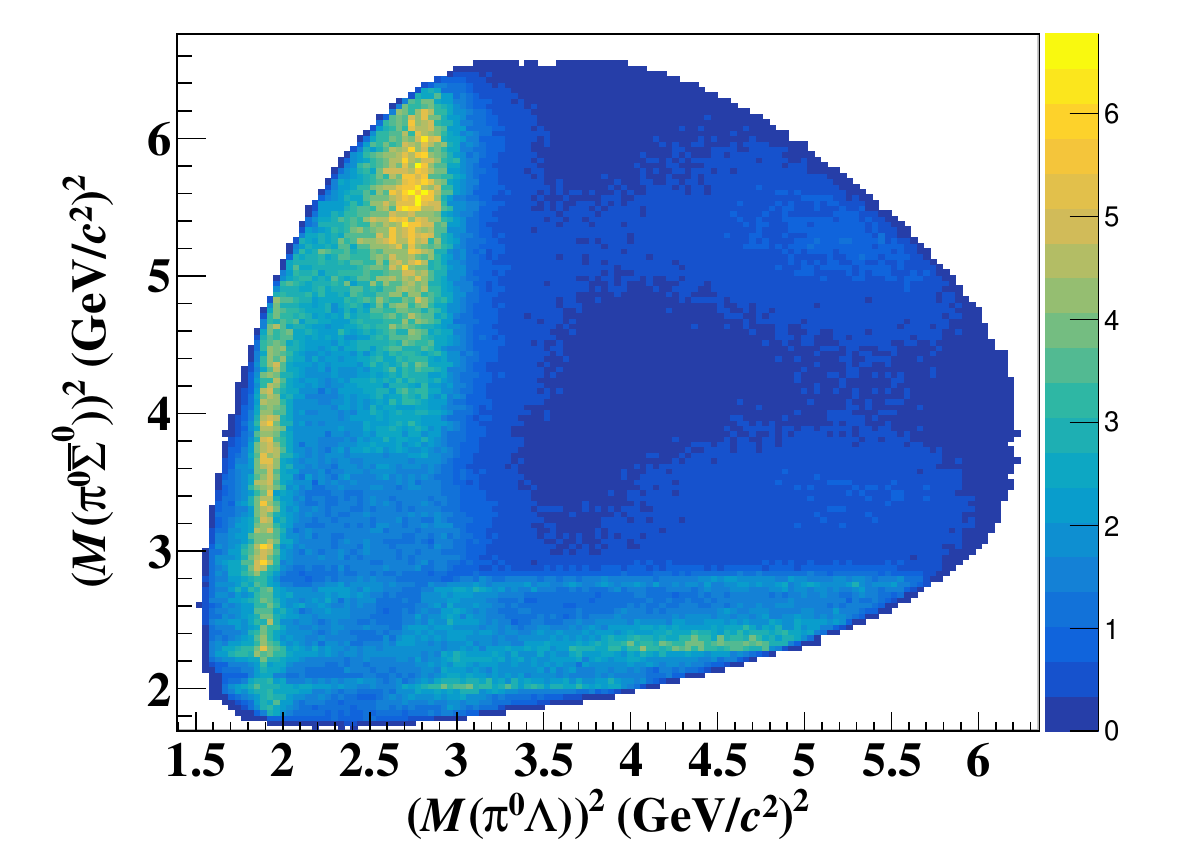}
    \caption{The Dalitz plots of $(M(\piz\bar{\Sigma}^0))^2\;\mathrm{versus}\;(M(\piz\Lambda))^2$ of data (left) and the PWA fit result (right).}
    \label{fig:pwa_dalitz}
\end{figure}

\begin{figure}[tbp]
    \centering
    \includegraphics[width=0.485\textwidth]{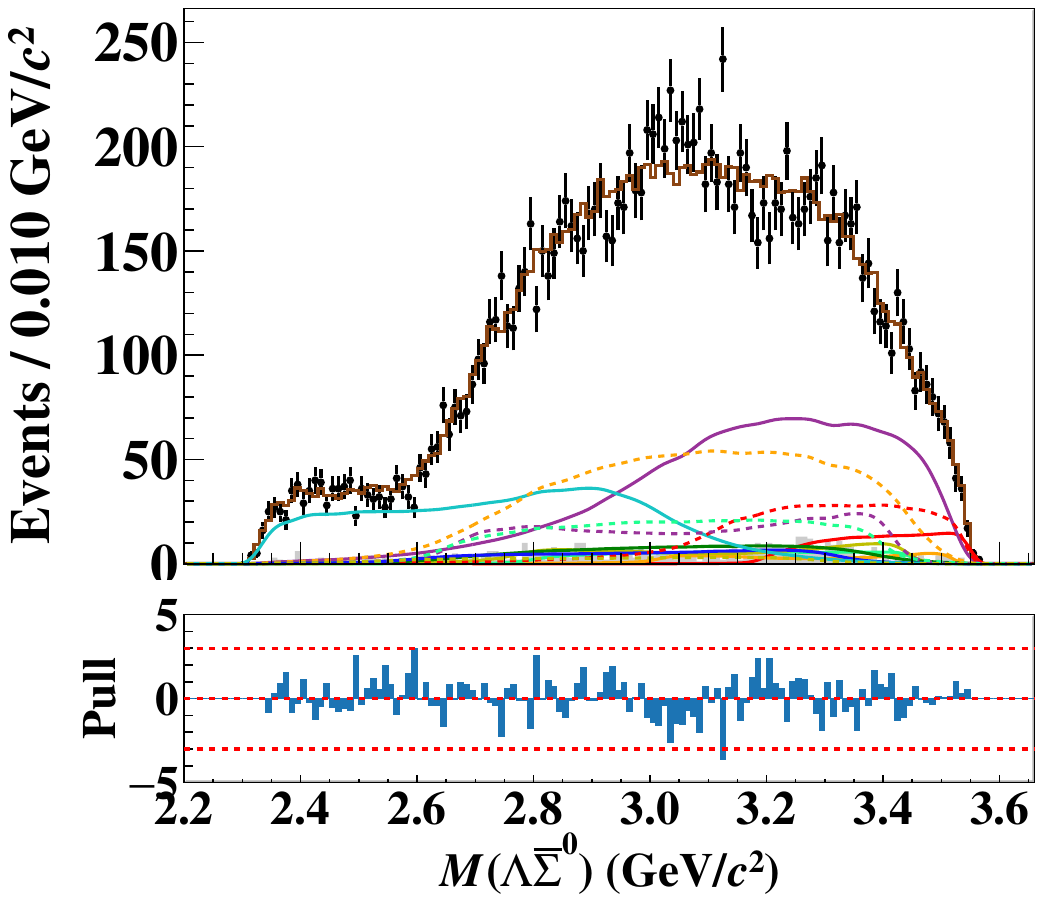}
    \includegraphics[width=0.485\textwidth]{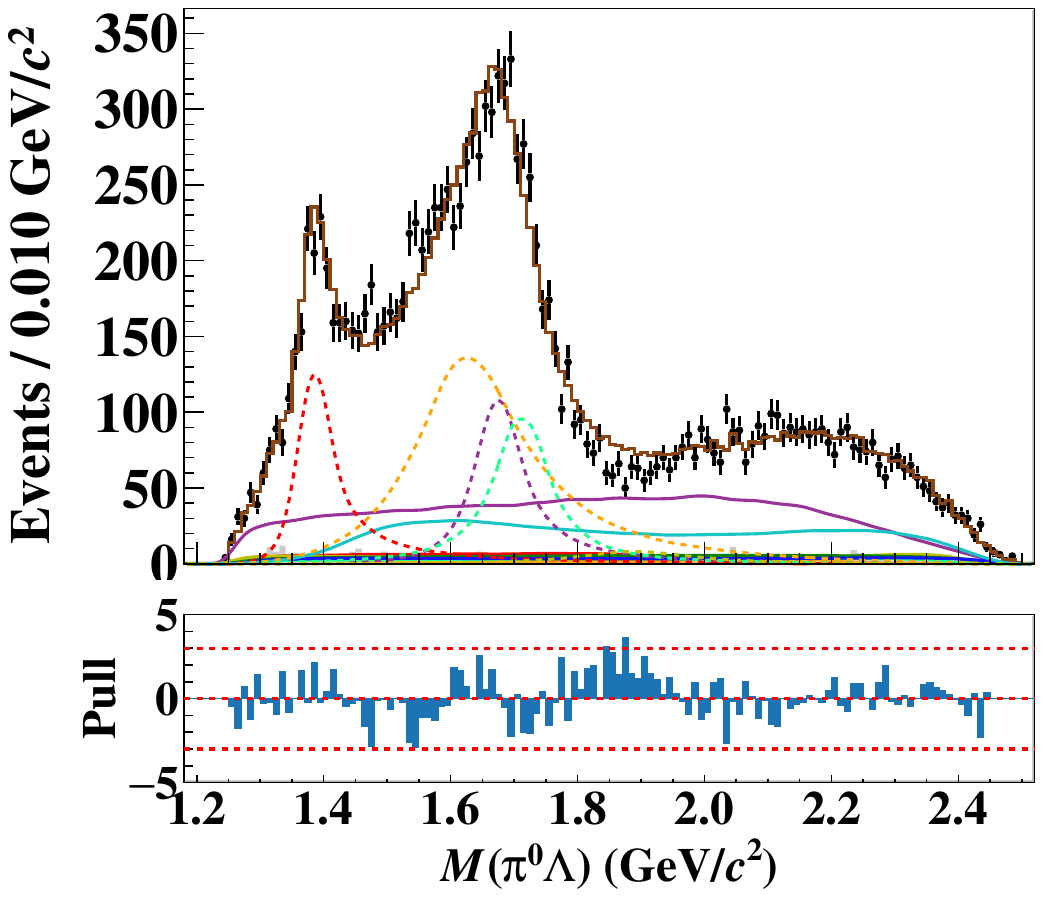}
    \includegraphics[width=0.485\textwidth]{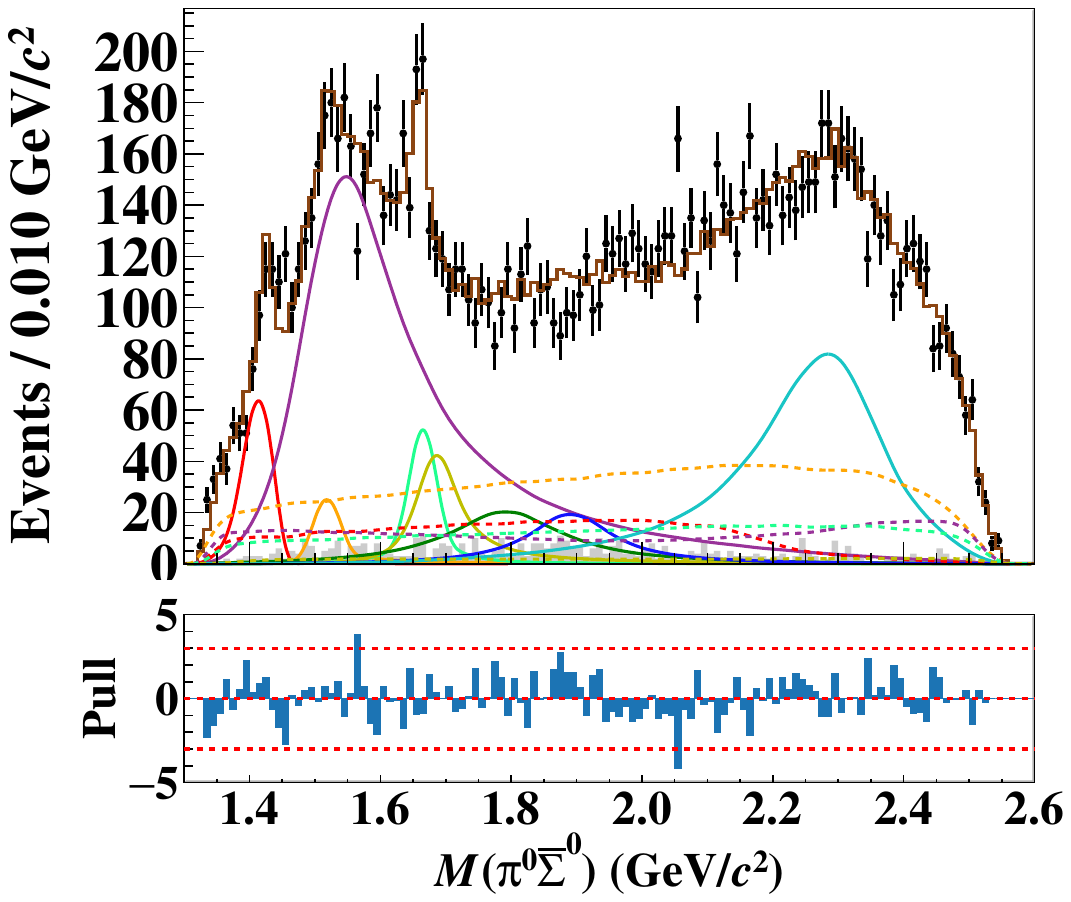}
    \includegraphics[width=0.485\textwidth]{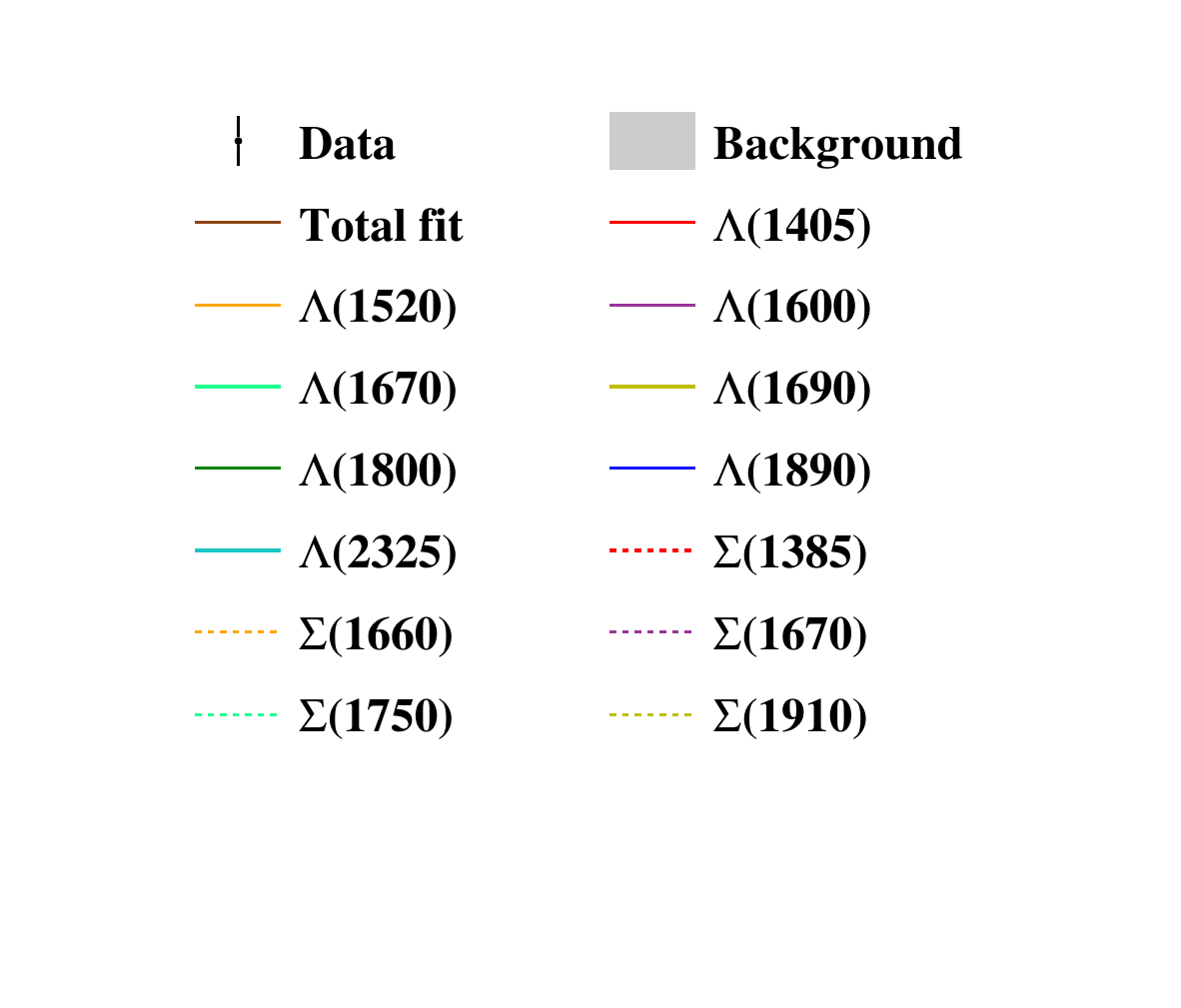}
    \caption{The distributions of $M(\LSb)$, $M(\piz\Lambda)$, and $M(\piz\bar{\Sigma}^0)$.}
    \label{fig:pwa_mass}
\end{figure}

\begin{figure}[tbp]
    \centering
    \includegraphics[width=0.485\textwidth]{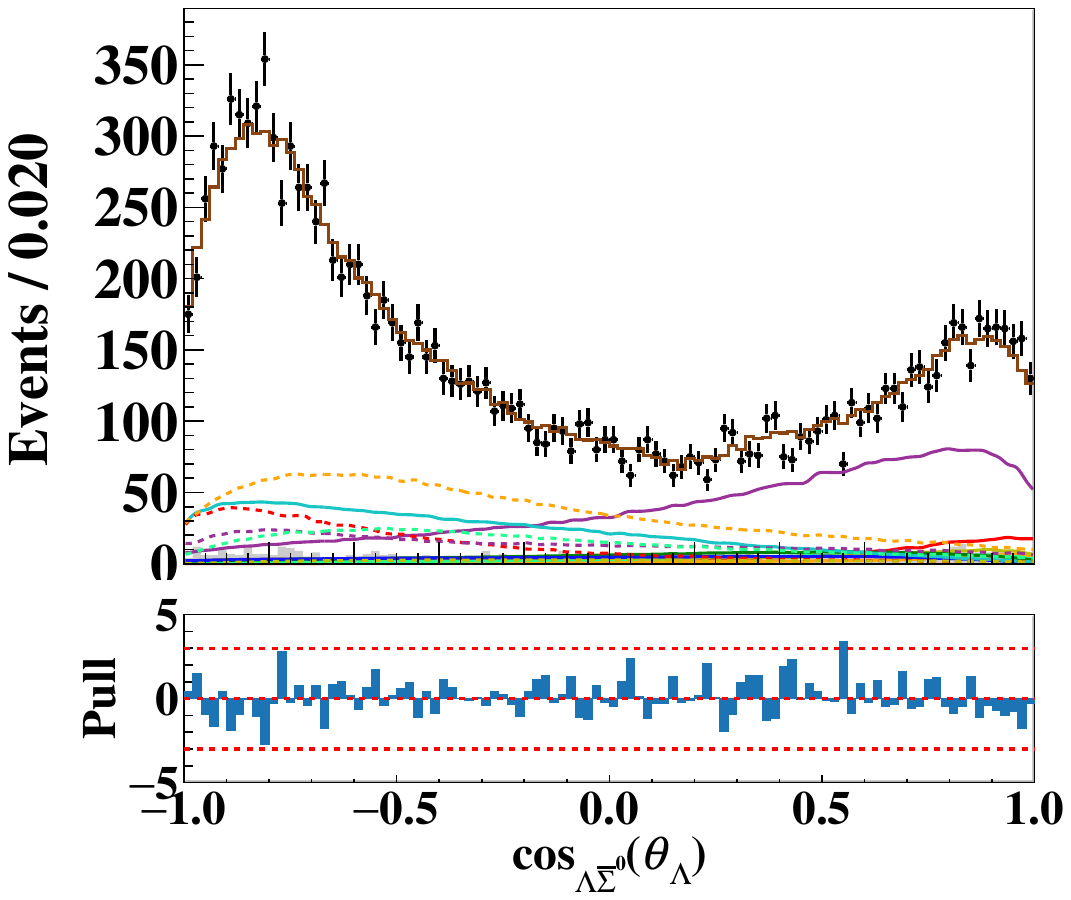}
    \includegraphics[width=0.485\textwidth]{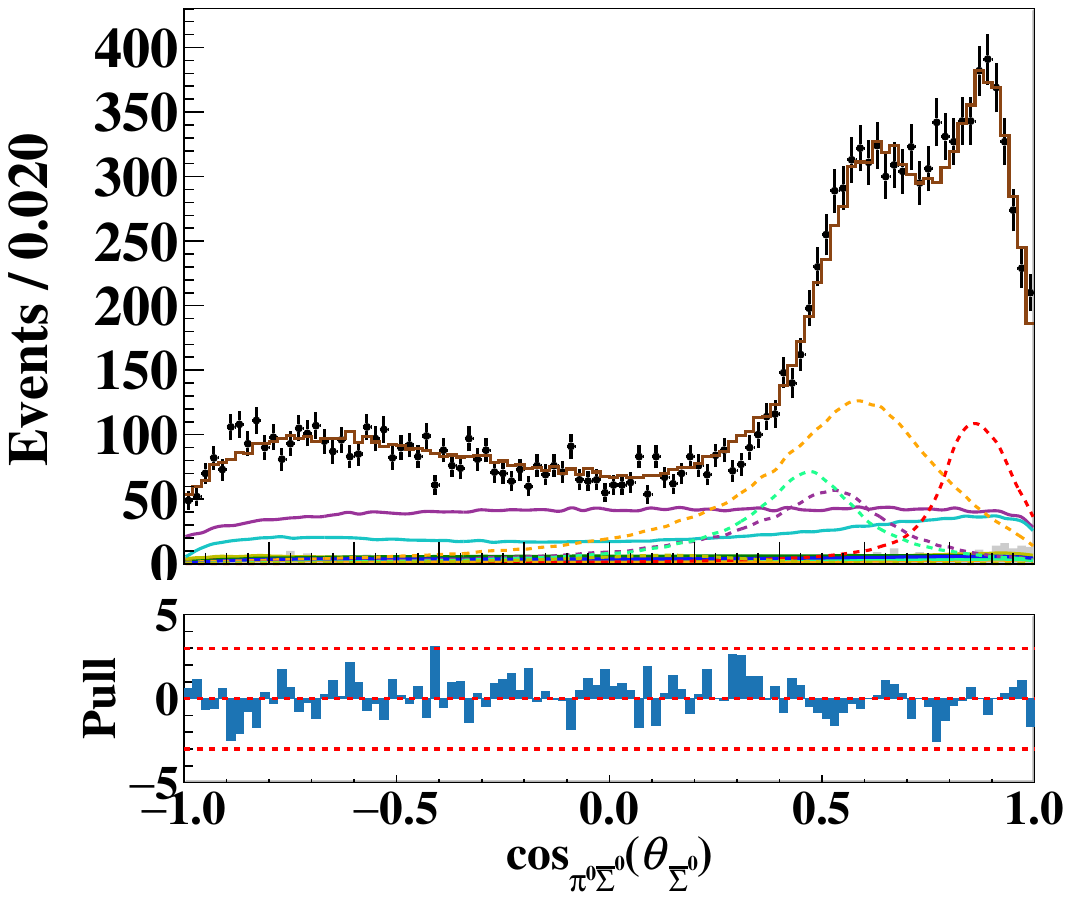}
    \includegraphics[width=0.485\textwidth]{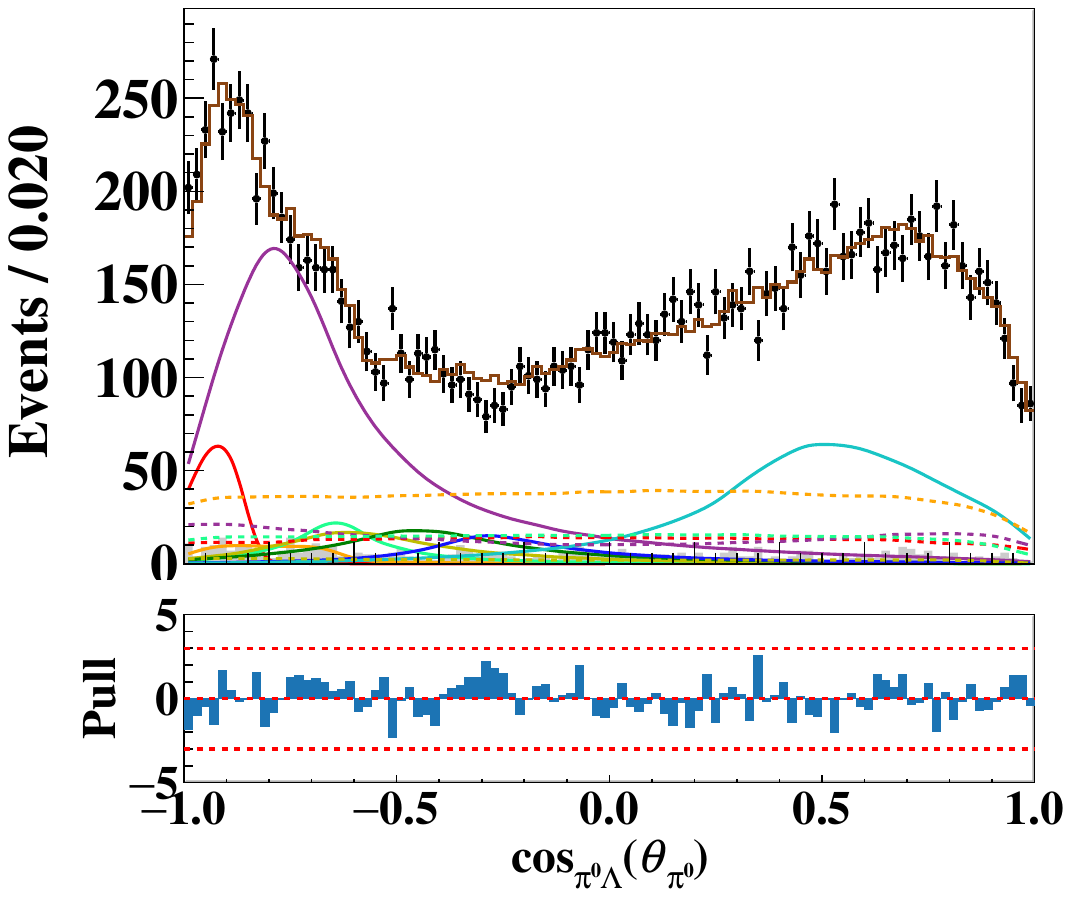}
    \includegraphics[width=0.485\textwidth]{Legend.pdf}
    \caption{The distributions of $\cos_{\LSb}(\theta_{\Lambda})$, $\cos_{\piz\bar{\Sigma}^0}(\theta_{\bar{\Sigma}^0})$, and $\cos_{\piz\Lambda}(\theta_{\piz})$.}
    \label{fig:pwa_angle}
\end{figure}

The obtained masses, widths and branching fractions of each resonance are listed in table~\ref{tab:pwa_result}, where the first uncertainty is statistical and the second systematic, the systematic uncertainty estimation is detailed in section~\ref{sec:syu}. The fit fractions and phases of each resonance are shown in table~\ref{tab:pwa_result_ff} with the statistical uncertainties only. The Dalitz plots of $(M(\piz\bar{\Sigma}^0))^2\;\mathrm{versus}\;(M(\piz\Lambda))^2$ of data and the PWA fit result are shown in figure~\ref{fig:pwa_dalitz}. The invariant mass distributions of $\LSb$, $\piz\Lambda$, and $\piz\bar{\Sigma}^0$ are shown in figure~\ref{fig:pwa_mass}. The helicity angular distributions are shown in figure~\ref{fig:pwa_angle}. From the $\piz\Lambda$ mass spectrum, it is observed that the fit quality is suboptimal in the region of $1.8\sim1.95\;\mathrm{GeV}/c^2$, motivating theoretical studies into the physical soundness of the lineshapes selected for resonant baryonic decays.

\begin{table}[tbp]
    \centering
    \begin{tabular}{c r@{$\;\pm\;$}l r@{$\;\pm\;$}l}
        \toprule
        Resonance       & \multicolumn{2}{c}{Fit fraction $(\%)$} & \multicolumn{2}{c}{Phase (rad)}                                            \\
        \midrule
        $\Lambda(1405)$ & $2.9$                                   & $0.3$                           & \multicolumn{2}{c}{0.0 (fixed)}          \\
        $\Lambda(1520)$ & $1.2$                                   & $0.3$                           & $-1.16$                         & $0.48$ \\
        $\Lambda(1600)$ & $29.1$                                  & $1.6$                           & $3.79$                          & $0.18$ \\
        $\Lambda(1670)$ & $2.1$                                   & $0.5$                           & $6.36$                          & $0.28$ \\
        $\Lambda(1690)$ & $3.6$                                   & $0.7$                           & $-8.79$                         & $0.34$ \\
        $\Lambda(1800)$ & $3.9$                                   & $1.0$                           & $-1.65$                         & $0.36$ \\
        $\Lambda(1890)$ & $3.1$                                   & $0.7$                           & $0.47$                          & $0.33$ \\
        $\Lambda(2325)$ & $18.9$                                  & $1.3$                           & $1.61$                          & $0.23$ \\
        $\Sigma(1385)$  & $8.4$                                   & $0.7$                           & $5.24$                          & $0.26$ \\
        $\Sigma(1660)$  & $22.9$                                  & $1.8$                           & $9.14$                          & $0.23$ \\
        $\Sigma(1670)$  & $10.3$                                  & $1.6$                           & $6.62$                          & $0.29$ \\
        $\Sigma(1750)$  & $9.0$                                   & $1.5$                           & $3.73$                          & $0.21$ \\
        $\Sigma(1910)$  & $1.5$                                   & $0.6$                           & $0.27$                          & $0.75$ \\
        \bottomrule
    \end{tabular}
    \caption{The fit fractions and phases of each component in the nominal PWA fit. The uncertainties are statistical only.}
    \label{tab:pwa_result_ff}
\end{table}

In chiral SU(3) dynamics, the existence of two poles in the $\Lambda(1405)$ region was first reported in ref.~\cite{Oller:2000fj}. This indicates that the $\Lambda(1405)$ resonance is not a single state but is instead a superposition of two eigenstates. The two-pole structure has a significant impact on hadron spectroscopy since it is related to the number of eigenstates in this sector. In general, a resonance state can have different coupling strengths to various channels. It was shown in ref.~\cite{Jido:2003cb} that the high-mass pole of the $\Lambda(1405)$ strongly couples to $\bar{K}N$ and the low-mass pole has a large coupling to $\pi\Sigma$, as expected from their origin in the isospin basis.

To test the stability of the chiral dynamics model, an alternative fit is performed using a Flatt\'{e}-like formula~\cite{Chung:1995dx} in place of the chiral dynamics model. The Flatt\'{e}-like formula, as shown in eq.~\eqref{eq:bw}, is operated by coupling the $\pi\Sigma$ and $\bar{K}N$ channels, where the masses of the $\Lambda(1405)$ and the final state particles are quoted from the PDG~\cite{ParticleDataGroup:2022pth}. The $\Delta NLL$, $\Delta N_\mathrm{dof}$, statistical significance and fit fraction of the $\Lambda(1405)$ are $68.0$, $4$, $11.1\sigma$ and $(3.0\pm0.3)\%$ in nominal fit, or $73.1$, $6$, $11.2\sigma$ and $(6.7\pm0.9)\%$ in alternative fit with the Flatt\'{e}-like formula, respectively. Due to the similar statistical significance of the $\Lambda(1405)$ with the two different parameterizations and the chiral dynamics model being well established in the literature, the parameterization of the chiral dynamics is found to be reasonable and is used in the nominal PWA fit.

\section{Summary}

Based on a sample of $(2712.4\pm14.3)\times10^6\;\psi(3686)$ events~\cite{BESIII:2024lks} collected with the BESIII detector, a PWA of $\psip\to\LSb\piz$ is performed for the first time to investigate possible $\Lambda^*$ and $\Sigma^*$ states. The measured masses, widths, and branching fractions for each component are summarized in table~\ref{tab:pwa_result}. In addition to the well established $\Lambda^*$ and $\Sigma^*$ states, including the $\Lambda(1405)$, $\Lambda(1520)$, $\Lambda(1600)$, $\Lambda(1670)$, $\Lambda(1690)$, $\Lambda(1800)$, $\Lambda(1890)$, $\Sigma(1385)$, $\Sigma(1660)$, $\Sigma(1670)$, $\Sigma(1750)$, and $\Sigma(1910)$, the PWA results indicate that the $\Lambda(2325)$ is necessary to better describe data. It is found that both the chiral dynamics model and the Flatt\'{e}-like formula for the parameterization of the $\Lambda(1405)$ can describe the data well. Due to the low statistics of the data, the two models are not distinguishable. Although the $\Lambda(2325)$ is a state of one-star of existence in the PDG~\cite{ParticleDataGroup:2022pth}, it is also one of the dominant contributions in $\psip\to\Lambda\bar{\Sigma}^0\piz$, which provides further proof of the existence of the $\Lambda(2325)$. We also perform a check on its spin-parity and find that its spin-parity favors $J^P=3/2^-$, which is consistent with that in the PDG~\cite{ParticleDataGroup:2022pth}. In addition, the branching fraction of $\psip\to\LSb\piz+c.c.$ is measured to be $(1.544\pm0.013\pm0.071)\times10^{-4}$ for the first time.

\begin{table}[tbp]
    \centering
    \begin{tabular}{c r@{$\;\pm\;$}c@{$\;\pm\;$}l r@{$\;\pm\;$}c@{$\;\pm\;$}l r@{$\;\pm\;$}c@{$\;\pm\;$}l}
        \toprule
        Resonance       & \multicolumn{3}{c}{Mass $(\mathrm{MeV}/c^2)$} & \multicolumn{3}{c}{Width ($\mathrm{MeV}$)} & \multicolumn{3}{c}{$\mathcal{B}\;(\times10^{-5})$}                                                        \\
        \midrule
        $\Lambda(1405)$ & \multicolumn{3}{c}{...}                       & \multicolumn{3}{c}{...}                    & $0.44$                                             & $0.05$  & $0.12$                                     \\
        $\Lambda(1520)$ & $1519.9$                                      & $1.6$                                      & $4.8$                                              & $20.6$  & $1.9$  & $0.6$  & $0.19$ & $0.05$ & $0.05$ \\
        $\Lambda(1600)$ & $1570.5$                                      & $4.6$                                      & $12.1$                                             & $228.1$ & $11.9$ & $34.8$ & $4.49$ & $0.25$ & $1.23$ \\
        $\Lambda(1670)$ & $1667.5$                                      & $2.3$                                      & $3.6$                                              & $30.2$  & $4.2$  & $8.7$  & $0.33$ & $0.07$ & $0.09$ \\
        $\Lambda(1690)$ & $1691.1$                                      & $4.4$                                      & $15.7$                                             & $72.3$  & $4.7$  & $18.3$ & $0.55$ & $0.11$ & $0.10$ \\
        $\Lambda(1800)$ & $1800.9$                                      & $13.3$                                     & $12.6$                                             & $208.8$ & $14.5$ & $36.7$ & $0.60$ & $0.15$ & $0.18$ \\
        $\Lambda(1890)$ & $1897.2$                                      & $9.6$                                      & $6.2$                                              & $149.2$ & $13.5$ & $4.1$  & $0.47$ & $0.10$ & $0.09$ \\
        $\Lambda(2325)$ & $2306.5$                                      & $6.3$                                      & $17.1$                                             & $227.1$ & $12.2$ & $47.8$ & $2.93$ & $0.20$ & $0.26$ \\
        $\Sigma(1385)$  & $1388.2$                                      & $1.9$                                      & $3.3$                                              & $60.5$  & $3.6$  & $6.3$  & $1.30$ & $0.11$ & $0.19$ \\
        $\Sigma(1660)$  & $1643.2$                                      & $4.5$                                      & $7.6$                                              & $221.3$ & $13.1$ & $41.1$ & $3.53$ & $0.29$ & $0.63$ \\
        $\Sigma(1670)$  & $1679.7$                                      & $3.4$                                      & $4.3$                                              & $87.0$  & $6.4$  & $8.9$  & $1.60$ & $0.25$ & $0.38$ \\
        $\Sigma(1750)$  & $1714.9$                                      & $4.2$                                      & $7.3$                                              & $97.2$  & $9.8$  & $9.7$  & $1.39$ & $0.23$ & $0.32$ \\
        $\Sigma(1910)$  & $1912.1$                                      & $10.6$                                     & $33.6$                                             & $225.1$ & $24.5$ & $46.0$ & $0.23$ & $0.10$ & $0.06$ \\
        \bottomrule
    \end{tabular}
    \caption{The masses, widths, and branching fractions of each component in the nominal PWA fit. The first uncertainties are statistical and the second systematic.}
    \label{tab:pwa_result}
\end{table}

\acknowledgments
The BESIII Collaboration thanks the staff of BEPCII and the IHEP computing center for their strong support. This work is supported in part by National Natural Science Foundation of China (NSFC) under Contracts Nos.
12075107, 12075250, 12075252, 12175244, 12247101, 12465015
11635010, 11735014, 11935015, 11935016, 11935018, 12025502, 12035009, 12035013, 12061131003, 12192260, 12192261, 12192262, 12192263, 12192264, 12192265, 12221005, 12225509, 12235017, 12361141819; 
National Key R\&D Program of China under Contracts Nos. 2020YFA0406300, 2020YFA0406400, 2023YFA1606000; 
the Fundamental Research Funds for the Central Universities under Contracts No. lzujbky-2024-jdzx06;
the Natural Science Foundation of Gansu Province under Contracts No. 22JR5RA389; 
the ‘111 Center’ under Contracts No. B20063; 
the Chinese Academy of Sciences (CAS) Large-Scale Scientific Facility Program; the CAS Center for Excellence in Particle Physics (CCEPP); Joint Large-Scale Scientific Facility Funds of the NSFC and CAS under Contract Nos. U2032110, U1832207; 100 Talents Program of CAS; The Institute of Nuclear and Particle Physics (INPAC) and Shanghai Key Laboratory for Particle Physics and Cosmology; German Research Foundation DFG under Contracts Nos. 455635585, FOR5327, GRK 2149; Istituto Nazionale di Fisica Nucleare, Italy; Ministry of Development of Turkey under Contract No. DPT2006K-120470; National Research Foundation of Korea under Contract No. NRF-2022R1A2C1092335; National Science and Technology fund of Mongolia; National Science Research and Innovation Fund (NSRF) via the Program Management Unit for Human Resources \& Institutional Development, Research and Innovation of Thailand under Contract No. B16F640076; Polish National Science Centre under Contract No. 2019/35/O/ST2/02907; The Swedish Research Council; U. S. Department of Energy under Contract No. DE-FG02-05ER41374

\newpage
{\bf \noindent The BESIII collaboration}\\
\\
{\small
M.~Ablikim$^{1}$, M.~N.~Achasov$^{4,c}$, P.~Adlarson$^{76}$, O.~Afedulidis$^{3}$, X.~C.~Ai$^{81}$, R.~Aliberti$^{35}$, A.~Amoroso$^{75A,75C}$, Q.~An$^{72,58,a}$, Y.~Bai$^{57}$, O.~Bakina$^{36}$
% ...
\\
{\it
$^{1}$ Institute of High Energy Physics, Beijing 100049, People's Republic of China\\
% ...
$^{3}$ Bochum  Ruhr-University, D-44780 Bochum, Germany\\
$^{4}$ Budker Institute of Nuclear Physics SB RAS (BINP), Novosibirsk 630090, Russia\\
% ...
$^{35}$ Johannes Gutenberg University of Mainz, Johann-Joachim-Becher-Weg 45, D-55099 Mainz, Germany\\
$^{36}$ Joint Institute for Nuclear Research, 141980 Dubna, Moscow region, Russia\\
% ...
$^{57}$ Southeast University, Nanjing 211100, People's Republic of China\\
$^{58}$ State Key Laboratory of Particle Detection and Electronics, Beijing 100049, Hefei 230026, People's Republic of China\\
% ...
$^{72}$ University of Science and Technology of China, Hefei 230026, People's Republic of China\\
% ...
$^{75}$ University of Turin and INFN, (A)University of Turin, I-10125, Turin, Italy; (B)University of Eastern Piedmont, I-15121, Alessandria, Italy; (C)INFN, I-10125, Turin, Italy\\
$^{76}$ Uppsala University, Box 516, SE-75120 Uppsala, Sweden\\
% ...
$^{81}$ Zhengzhou University, Zhengzhou 450001, People's Republic of China\\

\vspace{0.2cm}
$^{a}$ Deceased\\
% ...
$^{c}$ Also at the Novosibirsk State University, Novosibirsk, 630090, Russia\\
% ...
}}
\end{document}